# Stochastic inversion of Gaussian random media using transverse coherence functions for reflected waves

Hao Hu, Yingcai Zheng

The University of Houston, Department of Earth and Atmospheric Sciences,

hhu5@central.uh.edu; yzheng24@central.uh.edu

# 1 Abstract

The transverse coherence functions (TCFs) of phase and amplitude fluctuations of a seismic wave are powerful to estimate the spatial distribution, length scales, and strength of random heterogeneities. However, TCFs have been formulated for transmitted waves only, not for reflected waves. In this paper, we derive reflection TCFs for Gaussian random media. Furthermore, we propose to invert for Gaussian random media using the reflection TCFs based on the grid search. We validate the new reflection TCF formulas using 2D finite-difference numerical experiments. The numerical example also illustrates the feasibility and efficiency of the inversion. The stochastic inversion using reflected waves can be used in both exploration and global seismology.

# 2 Introduction

Earth is heterogeneous across multiple scales and the heterogeneities may be due to variation in rock composition, porosity, fluid content, or thermal states. Information of random heterogeneities in the global Earth can be used to infer dynamics and mixing processes (e.g., Anderson, 2006; Xu *et al.*, 2008; Li and Zheng, 2019). A full deterministic description of the heterogeneities is neither possible nor desirable and a statistical description can be much more practical and useful. In exploration geophysics, the knowledge of small-scale heterogeneities is also important to assess the oil/gas volume in the reservoir evaluations in fossil resource exploration and production (e.g., Huang *et al.*, 2012; Meng *et al.*, 2017).

Many seismic scattering methods have been developed to characterize random media in terms of their statistical parameters in different regions of the Earth. Examples include well-logging analysis (e.g., Wu *et al.*, 1994; Sivaji *et al.*, 2002; Fukushima *et al.*, 2003), seismic tomography (e.g., Meschede and Romanowicz, 2015; Nakata and Beroza, 2015), seismic envelope analysis for the mantle (e.g., Aki and Chouet, 1975; Wu, 1982; Sato, 1984; Jannaud *et al.*, 1991; Hedlin and Shearer, 2002; Emoto *et al.*, 2017) and the core (e.g., Vidale and Earle, 2000; 2005; Peng *et al.*, 2008). The transmission fluctuation analysis using seismic array data is another method and is the focus of our paper (e.g., Flatte and Wu, 1988; Wu and Flatte, 1990; Chen and Aki, 1991; Sivaji *et al.*, 2001; Zheng and Wu, 2005; 2008; Zheng, 2012; Yoshimoto *et al.*, 2015; Cormier *et al.*, 2018).

In geophysics, Aki (1973) pioneered to infer to the spatial spectrum of velocity heterogeneity from measuring seismic wave amplitude and phase fluctuations using the Chernov theory (Chernov, 1960). Flatte and Wu (1988) developed the angular coherence functions (ACFs), to estimate the depth-dependent heterogeneity spectra. Wu and Flatte (1990) further formulated the ACFs and TCFs to explain observations with more complicated heterogeneities. Later Wu and Xie (1991) developed the joint transverse-angular coherence functions (JTACF) to invert for the heterogeneity spectra with a better depth resolution. Chen and Aki (1991) independently derived the JTACF, but using the Born approximation that is different from Wu's works (Wu and Flatte, 1990; Wu and Xie, 1991) based on the Rytov and parabolic approximations. Line *et al.* (1998) tried to use reflected waves from active sources in the stochastic inversion of TCFs but still based on the transmission TCF formulas. Zheng and Wu (2005) discussed the measurements of phase and amplitude and pointed out that the unwrapped phase and spectral amplitude should be used in forming TCFs. Zheng *et al.* (2007) proposed to invert for the heterogeneity spectrum of a single layer of stationary random heterogeneities via the Fourier transform of the sum of the log-amplitude and phase TCFs. Zheng and Wu (2008) further extended the TCF theory for a depth-dependent background media from the traditional assumption of constant background media.

However, commonly used TCF methods were formulated for transmitted waves. The reflection case has not been done analytically. Transmission TCFs are useful for teleseismic or cross-well observations. The reflection TCFs will be useful not only in earthquake seismology but also for exploration seismology where surface seismic surveys are readily available due to active sources.

In this paper, we first review the TCF method and propose to the TCFs of reflected waves for a heterogeneous layer of finite thickness, containing Gaussian random heterogeneities. We then validate the accuracy of reflection TCFs using numerical experiments and finite-difference acoustic wave simulations. Finally, we illustrate how to invert for the Gaussian heterogeneity spectrum using reflection TCFs.

# 3 Reflection transverse coherence functions (TCFs) theory in random media

Here, we formulate the reflection TCFs. We consider the reflection geometry shown in Figure 1 and a plane wave is incident upon a random medium layer and then reflected up to the receivers, by a reflector at the bottom of the random layer (Figure 1).

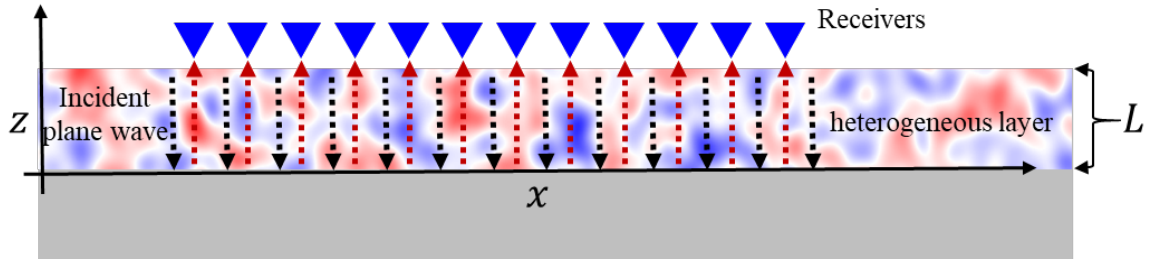

Figure 1. The schematic geometry of receivers (blue triangles) and a heterogeneous layer with Gaussian random perturbations. The thickness of the heterogeneous layer is $L$. With an incident plane wave (downward black-dashed arrows), receivers will record waves (upward red-dashed arrows) reflected from the lower-boundary of the heterogeneous layer.

We measure the logarithmic amplitude (logA), $u(x)$, and phase, $\phi(x)$, of the reflected wave, recorded at the surface receiver located at $x$. Usually, the measurement is done at a particular frequency. The amplitude is the spectral amplitude and the phase is the unwrapped phase (Zheng and Wu, 2005), for the reflected arrival. The TCFs for the logA ($\langle uu \rangle$) and the phase ($\langle \phi\phi \rangle$) are defined as:

$$\langle uu \rangle = \langle u(x_1)u(x_2) \rangle,$$

$$\langle \phi\phi \rangle = \langle \phi(x_1)\phi(x_2) \rangle.$$

(1)

⟨ ⟩ means the ensemble averaging; $x_1$ and $x_2$ are two receiver locations. For a stationary and isotropic (i.e., no preferred directions in the statistical sense) random medium, both ⟨$uu$⟩ and ⟨$\phi\phi$⟩ depend only on the station transverse distance, $|x_2 - x_1|$.

The random medium is characterized by a stationary random velocity field, $v = v(x, z)$. We define the velocity perturbation as $\delta v(x, z) = \frac{1}{2}(\frac{v_0^2}{v^2} - 1)$; $v_0$ is a constant background velocity of the medium. The random medium has a correlation function,

$$W(|x'' - x'|, |z'' - z'|) = \langle \delta v(x'', z'')\, \delta v(x', z') \rangle, \tag{2}$$

where $(x', z')$ and $(x'', z'')$ indicates two arbitrary perturbation positions in the random medium. Because the random medium is stationary, $W$ only depends on the relative position of these two positions. In the stochastic random medium inversion, we invert for $W$, not the velocity field $v(x, z)$.

In the following, we use a "mirror reflection principle" to show how to derive the reflection TCF formula from the medium correlation function, $W$. We treat the reflector as a mirror, which creates a mirror image for both the random medium and the receivers (Figure 2). Now we have converted this problem into a transmission TCF problem.

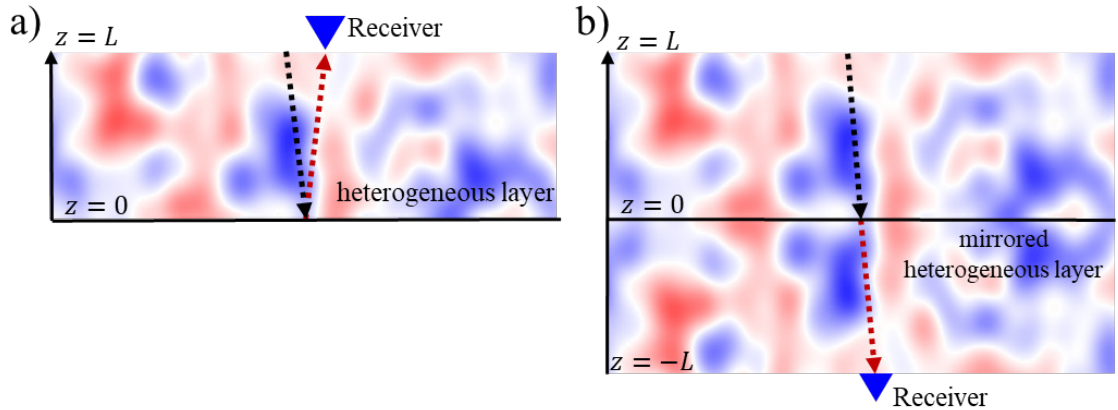

Figure 2. The geometry used in the derivation of reflection TCFs. (a) Reflected wave ray path in the random medium form the source to the receiver; (b) transmitted path in the mirror medium from the source to the mirror receiver. The thickness of the heterogeneous layer is $L$. In b), the mirrored heterogeneous layer is symmetric about $z = 0$.

## 3.1 Transmission TCFs

We first need to review the TCF theory for transmitted waves. Assume there is a Gaussian random heterogeneous layer with a total thickness of $H$. The incident wave is a downgoing plane wave. In a 2D acoustic wavefield, the amplitude and phase TCFs can be expressed as (Zheng *et al.*, 2007):

$$\langle uu \rangle = I_1(r_x) + I_2(r_x)$$

$$\langle \phi\phi \rangle = I_1(r_x) - I_2(r_x)$$

(3)

where $\langle \ \rangle$ implies the ensemble average; $\langle uu \rangle$ and $\langle \phi\phi \rangle$ are the logA and phase TCFs for a transverse lag between two receivers ($r_x = |x_2 - x_1|$), respectively; $I_1(r_x)$ and $I_2(r_x)$ are two auxiliary functions that can be obtained:

$$I_1(r_x) = \frac{k^2}{4\pi^2} \int d\kappa e^{i\kappa_x r_x} \int_0^H \widetilde{W}(\kappa_x, \eta) \cos\left(\frac{\eta \kappa_x^2}{2k}\right)(H - \eta) d\eta$$

$$I_2(r_x) = \frac{k^3}{4\pi^2} \int \frac{d\kappa_x}{\kappa_x^2} e^{i\kappa_x r_x} \int_0^H \widetilde{W}(\kappa_x, \eta) \left[\sin\left(\frac{\eta \kappa_x^2}{2k}\right) - \sin\left(\frac{2H - \eta}{2k}\kappa_x^2\right)\right] d\eta$$

(4)

where $k = \omega/v_0$ is the wavenumber of the measured frequency $\omega$ in the background velocity $v_0$;

$$\widetilde{W}(\kappa_x, \eta) = \frac{1}{2\pi} \int_{-\infty}^{\infty} e^{-i\kappa_x r_x} W(r_x, \eta) dr_x$$

(5)

is the medium perturbation correlation function $W(r_x, \eta)$ with correlation lags $|x'' - z'| = r_x$ and $|z'' - z'| = \eta$ at a lateral wavenumber of $\kappa_x$.

Additionally, the power spectrum density function (PSDF) $P(\kappa_x, \kappa_z)$ of the random medium could be estimated from the $W(r_x, \eta)$ by Fourier transform in an infinite medium (Tatarskii, 1971; Wu and Flatte, 1990):

$$P(\kappa_x, \kappa_z) \sim \frac{1}{4\pi^2} \int \int dr_x d\eta \, e^{-i\kappa_x r_x - i\kappa_z \eta} W(r_x, \eta) \quad (6)$$

To evaluate $I_1$ and $I_2$ integrals in TCFs (equation (4)), Zheng *et al.* (2007) assumed that $W(r_x, |z'' - z'|) \approx 0$ if $|z'' - z'| > \ell_z$ where $\ell_z$ is the vertical length scale of the heterogeneities and usually very small. In many cases, people just used the delta-correlation assumption, $W(r_x, |z'' - z'|) \propto W(r_x, 0) \delta(z'' - z')$.

## 3.2 Reflection TCFs

However, in our reflection TCFs case, we consider the TCF in a mirror medium (Figure 2b). In this case, the TCF formulas are the same except we need to consider, $W(r_x, |z'' - z'|)$ for $|z'' - (-z')| > \ell_z$ due to the mirror reflection of the random medium.

Because of the mirror, we need to modify $W$ in equation (5). From the derivation in Appendix A1, for a mirrored heterogeneous layer with isotropic Gaussian random perturbations, we can obtain the following correlation function and its PSDF in 2D:

$$W_{mirror}(r_x, r_z) = \varepsilon^2 \exp(-\tfrac{r_x^2}{a^2}) \left\{ \exp\left(-\tfrac{r_z^2}{a^2}\right) + \frac{1}{2L}\left[\tfrac{1}{2}\sqrt{\pi}a \, erf\left(\tfrac{r_z}{a}\right) - r_z \exp\left(-\tfrac{r_z^2}{a^2}\right)\right] \right\}, \quad (7)$$

$$P_{mirror}(\kappa_x, \kappa_z) = \varepsilon^2 \pi a^2 \exp\left(-\frac{1}{4}\kappa_x^2 a^2\right)\left[\exp\left(-\frac{1}{4}\kappa_z^2 a^2\right) + \frac{i}{4L}\left(a^2\kappa_z - \frac{2}{\kappa_z}\right)\right], \quad (8)$$

where $r_x$ and $r_z$ are the horizontal and vertical correlation lags; $\kappa_x$ and $\kappa_z$ are their corresponding wavenumbers; $\varepsilon$ and $a$ are the Gaussian perturbation strength and scale, respectively; $2L$ is the total thickness of the mirrored random medium (Figure 2b);

To compute the reflection TCFs, we just need to plug expressions (7) into equations 3-5 to replace $W$ and its spectrum $\widetilde{W}$.

## 4 Modeling TCFs for Gaussian random media

To validate the new formulas of reflection TCFs, we build a numerical model containing a Gaussian random layer (Figure 3) to model the reflected waves. The perturbation correlation functions follow the Gaussian function with a correlation length $a = 100$ m and a perturbation strength $\varepsilon = 0.01$. The thickness of the heterogeneous layer is 500 m (Figure 3a). We use a staggered-grid finite-difference (FD) numerical solution of the two-way acoustic wave equations to model the wave propagation in the numerical model (e.g., Virieux, 1986; Graves, 1996). The wavelet of the incident vertical plane wave is a Ricker wavelet with a dominating frequency at 20 Hz. The wavelength is 200 m. One example showing the recorded waveforms is in Figure 3b. The measured phase and logA at 20 Hz are in Figure 3c and d. After 100 random realizations, we can obtain the ensemble averages for phase and logA TCFs (in Figure 4). Meanwhile, in Figure 4 we can compare the measured TCFs from reflected waves and the theoretical reflection TCFs (equations (3),(4),(5), and (7)). Overall, the measured TCFs fit well with the theoretically predicted TCFs. It also validates our new formulas for the reflected waves.

There are some small discrepancies between the measured TCFs and theoretical TCFs due to the limited random realizations and model discretization in the numerical modeling of wave propagations. It can also be due to the use of the mirror assumption that neglects the interference of the backscattering effect because it uses transmission geometry for the reflected wave (Line *et al.*, 1998).

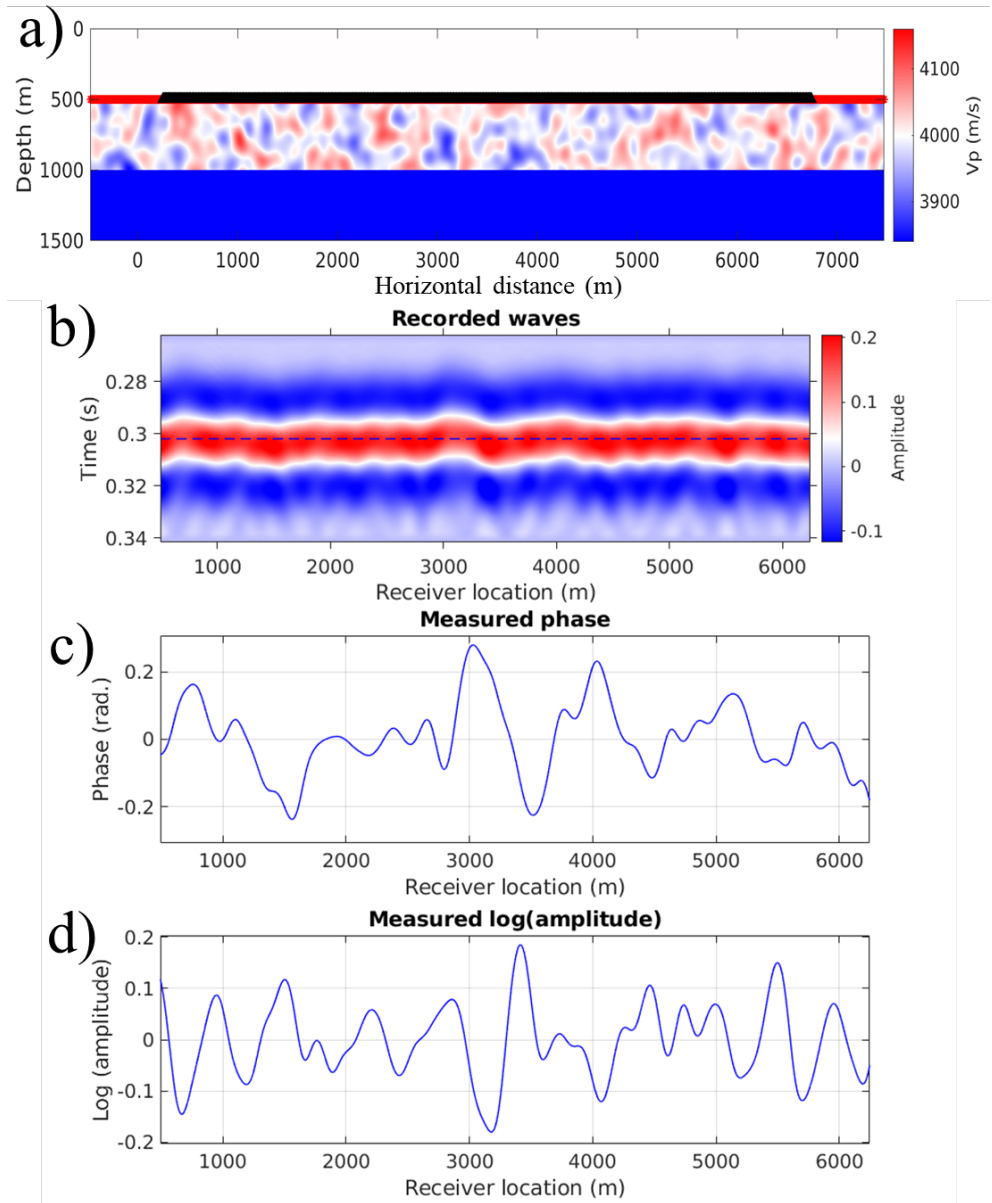

Figure 3. Numerical modeling of the reflected wave fluctuation. (a) The numerical model and geometry for modeling the wave. The incident wave is a vertical plane wave using vertical linear sources at depth of 500 m (red dots). The depth of receivers is 500 m (represented by black triangles). The receivers are horizontally deployed from 250 m to 7750 m at an interval of 5m. In the heterogeneous layer, the background $P$-velocity is 4000 $m/s$ and the density is $\rho = 2000\ g/cm^3$. The layer beneath the heterogeneous layer is of $v_p = 2000\ m/s$ and $\rho = 1000\ g/cm^3$ to generate strong reflected waves. The recorded reflected waves are in (b). (c) and (d) are the measurements of the phase and logA.

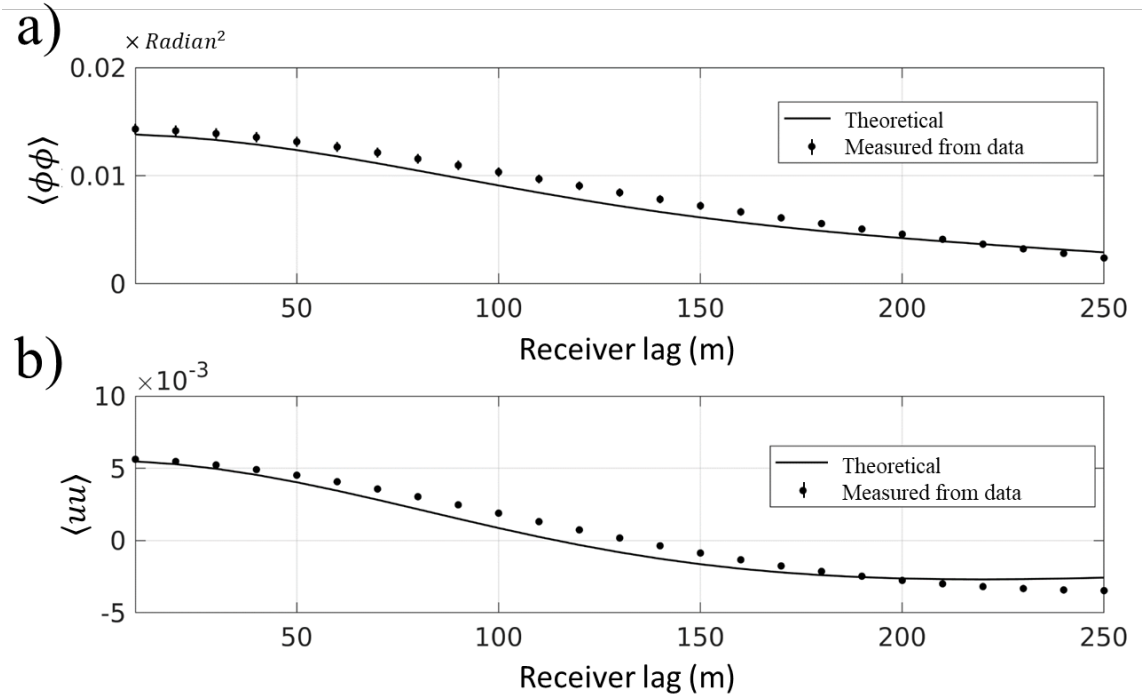

Figure 4. Comparison of measured TCFs and theoretical TCFs. (a) Phase TCFs; (b) logA TCFs. The back lines represent the theoretical TCFs for the reflected waves while the black dots with error bars (one standard deviation) are the measured TCFs for reflected waves. We use 100 realizations of bootstrap to obtain the averaged measurements and measurement errors from 100 numerical simulations.

# 5 Inversion of the heterogeneous medium using reflected waves

Once we have obtained the phase and logA TCF from data, we can grid search the Gaussian random medium parameters (perturbation strength $\varepsilon$ and scale $a$) to find out a pair of optimal parameters whose theoretical TCFs best fit the measured TCFs from observed data. We assume we know the random layer thickness, $L$. We can formulate this inverse problem by minimization of a misfit function:

$$F(r_o, \varepsilon) = |\langle uu \rangle(a, \varepsilon) - \langle uu \rangle_{data}| + |\langle \phi\phi \rangle(a, \varepsilon) - \langle \phi\phi \rangle_{data}|, \qquad (9)$$

where $\langle uu \rangle(a, \varepsilon)$ and $\langle \phi\phi \rangle(a, \varepsilon)$ are theoretical TCFs for a trial $a$ and $\varepsilon$, using equations (3), (4), (5), and (7). $\langle uu \rangle_{data}$ and $\langle \phi\phi \rangle_{data}$ are the TCFs measured from the recorded reflected waves. We will validate the inversion of Gaussian random medium parameters ($a$ and $\varepsilon$) using numerical experiments from the previous section.

Based on the measured TCFs of reflected waves (in Figure 4), we perform a grid search to calculate the misfit function between the measured reflection TCFs and the theoretical reflection TCFs using equation (9). The trial $a$ varies from 50 m to 200 m at an interval of 5 m while $\varepsilon$ from 0.05 to 0.015 at an interval of 0.005. The misfit map is shown in Figure 5 as a function of $\varepsilon$ and $a$. The global minimum occurs at $a = 110\ m$ and $\varepsilon = 0.095$, which are very close to the true values of the Gaussian random model. This numerical experiment illustrates that the inversion using reflection TCFs base on grid search is feasible and efficient.

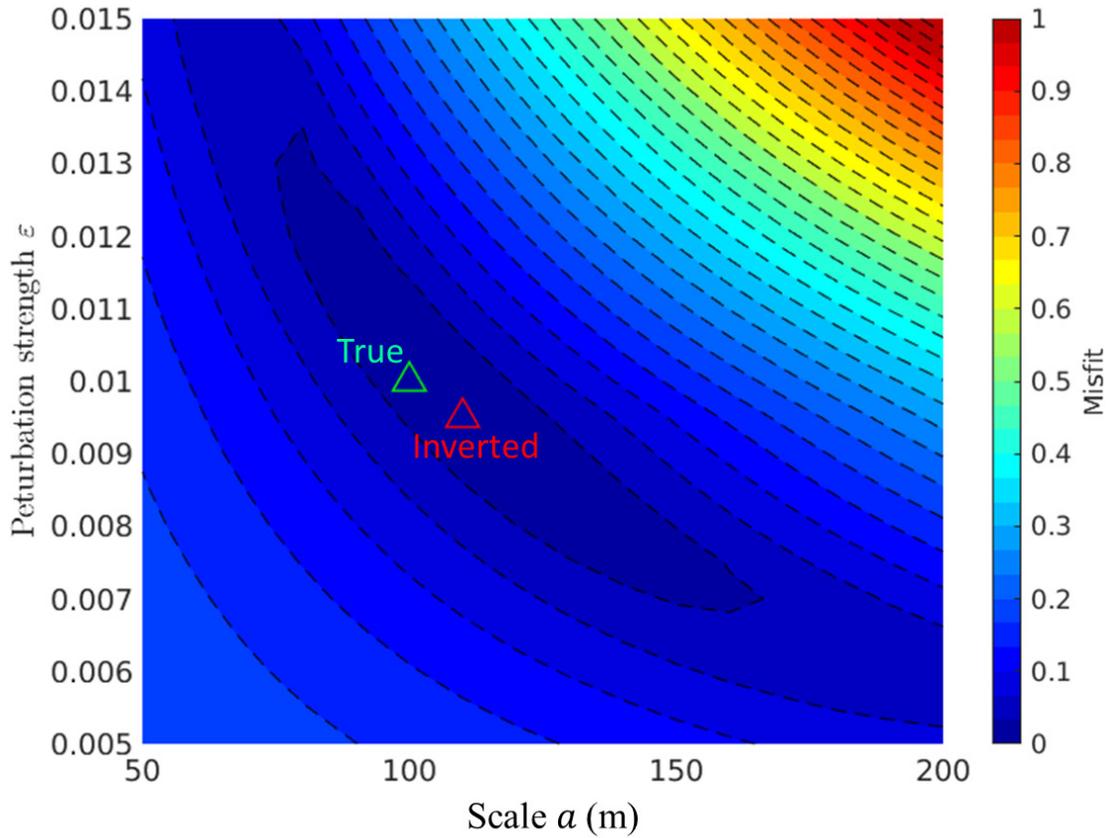

Figure 5. The normalized misfit map is a function of trial Gaussian correlation length $a$ and perturbation strength $\varepsilon$. The optimized solution of ($a = 110\ m$, $\varepsilon = 0.095$) is marked by the red triangle while the true values ($a = 100\ m$, $\varepsilon = 0.095$) is the green triangle.

# 6  Discussions

Line *et al.* (1998) tried to characterize the random heterogeneities in the crust using phase and logA measured from the reflection data. However, they used the transmission TCF formula to invert for the random medium parameters. Because the reflected path is two-way, in their TCF inversion the random layer thickness is doubled and they did not use the mirror technique we proposed here. Therefore, it would be worthwhile to investigate the error in using their proposed approach.

We consider two cases:

- Case 1 (Reflection). Same to Figure 3a, we use FD to model the reflected wave and measure the phase and logA fluctuations, then we form the TCFs (i.e., measured reflection TCFs). We have also used our proposed reflection TCF formulas to compute the theoretical reflection TCFs (Figure 6b and c). The receivers are on the top of the model and the incident plane wave is coming down from above.

- Case 2 (Transmission). We build a Gaussian random medium with the same statistical parameters ($a$ and $\varepsilon$) as in Case 1. However, the layer thickness is doubled. The receivers are placed at the bottom of the random layer and the incident plane wave is coming down (Figure 6a). We use FD to model the transmitted wave and measure its phase and logA fluctuations to form the TCFs (i.e., measured transmission TCFs). We also use the transmission TCF formula to compute the theoretical transmitted TCFs.

We can see in both Case 1 and Case 2, the measured and the theoretically predicted TCFs agree well with each case (Figure 6b and c). We can also see the phase TCF for the reflection case is overall larger than the transmission case, up to ~ 30%. Our proposed mirrow approach did a better job. The strength of the logA TCF for the reflection case is similar to the transmission case. These differences demonstrate that in order to invert for the random medium parameters using the measured reflection TCFs, it is necessary to use our newly derived formulas.

The newly proposed reflection TCF formula is derived for Gaussian random media. For non-Gaussian media, such as the von karman-type (e.g., Wu, 1982; Klimeš, 2002; Sato *et al.*, 2012; Sato, 2019), we can easily modify the random medium correlation function using the mirror reflection. Moreover, our formulas could be easily extended to 3D by simply changing the form of the perturbation correlation function.

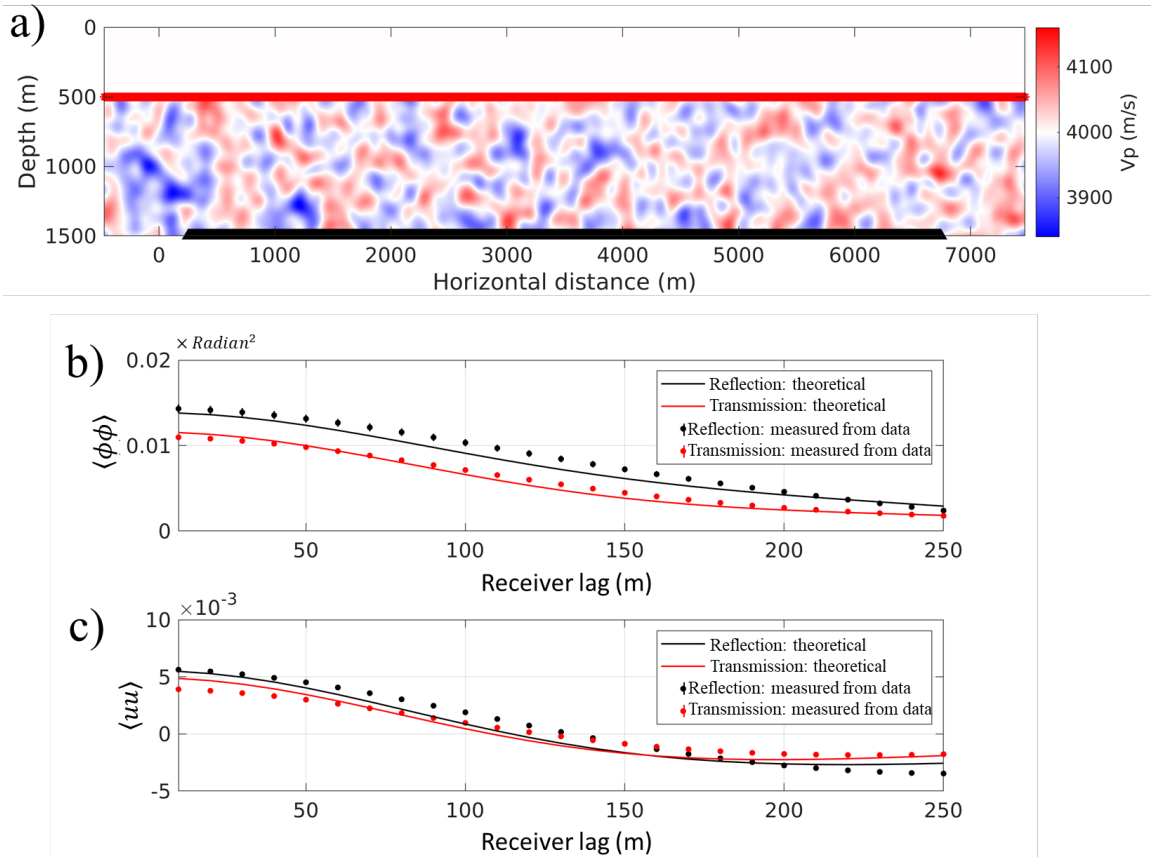

Figure 6. The comparison of the TCFs in Case 1 (reflection) and Case 2 (transmission). (a) Model and geometry for modeling transmitted waves in case 2. (b) and (c) are comparisons between phase TCF and logA TCF. In (b) and (c), the back lines represent the theoretical reflection TCFs while the red lines are theoretical transmission TCFs. The black dots with error bars (one standard deviation) are measured TCFs from reflected waves while red dots with error bars are from transmitted waves. We use 100 realizations of bootstrap to obtain the averaged measurements and measurement errors from 100 numerical simulations.

# 7 Conclusions

We derived new formula for reflection TCFs for the phase and logA fluctuations using a mirror reflection approximation. We validate the new formula using 2D numerical experiments via full-wave finite-difference modeling of wave propagation in random media. The measured phase and logA TCFs of reflected waves match well with those theoretical TCFs. We then showed how to use measured reflection TCF data to invert for Gaussian medium parameters, including the heterogeneity scale and the random perturbation strength using a grid-search scheme. Our newly derived formula and our stochastic inversion of TCFs using reflected waves have many potential applications in exploration geophysics for resource characterization and in global seismology.

Meschede, M., and B. Romanowicz (2015), Lateral heterogeneity scales in regional and global upper mantle shear velocity models, *Geophysical Journal International*, *200*(2), 1076-1093.

Nakata, N., and G. C. Beroza (2015), Stochastic characterization of mesoscale seismic velocity heterogeneity in long beach, california, *Geophysical Journal International*, *203*(3), 2049-2054.

Peng, Z. G., K. D. Koper, J. E. Vidale, F. Leyton, and P. Shearer (2008), Inner-core fine-scale structure from scattered waves recorded by LASA, *Journal of Geophysical Research-Solid Earth*, *113*(B9).

Sato, H. (1984), Attenuation and Envelope Formation of 3-Component Seismograms of Small Local Earthquakes in Randomly Inhomogeneous Lithosphere, *Journal of Geophysical Research*, *89*(Nb2), 1221-1241.

Sato, H. (2019), Power spectra of random heterogeneities in the solid earth, *Solid Earth*, *10*(1), 275-292.

Sato, H., M. C. Fehler, and T. Maeda (2012), *Seismic wave propagation and scattering in the heterogeneous earth*, Springer Science & Business Media.

Shapiro, S. A., and G. Kneib (1993), Seismic attenuation by scattering - theory and numerical results, *Geophysical Journal International*, *114*(2), 373-391.

Sivaji, C., O. Nishizawa, and Y. Fukushima (2001), Relationship between fluctuations of arrival time and energy of seismic waves and scale length of heterogeneity: An inference from experimental study, *Bulletin of the Seismological Society of America*, *91*(2), 292-303.

Sivaji, C., O. Nishizawa, G. Kitagawa, and Y. Fukushima (2002), A physical-model study of the statistics of seismic waveform fluctuations in random heterogeneous media, *Geophysical Journal International*, *148*(3), 575-595.

Tatarskii, V. I. (1971), *The effects of the turbulent atmosphere on wave propagation*.

Vidale, J. E., and P. S. Earle (2000), Fine-scale heterogeneity in the Earth's inner core, *Nature*, *404*(6775), 273-275.

Vidale, J. E., and P. S. Earle (2005), Evidence for inner-core rotation from possible changes with time in PKP coda, *Geophysical Research Letters*, *32*(1).

Virieux, J. (1986), P-SV-WAVE PROPAGATION IN HETEROGENEOUS MEDIA - VELOCITY-STRESS FINITE-DIFFERENCE METHOD, *Geophysics*, *51*(4), 889-901.

Wu, R. S. (1982), Attenuation of short period seismic waves due to scattering, *Geophys. Res. Lett.*, *9*.

Wu, R. S., and S. M. Flatte (1990), Transmission fluctuations across an array and heterogeneities in the crust and upper mantle, *Pure and Applied Geophysics*, *132*(1-2), 175-196.

Wu, R. S., and X. B. Xie (1991), Numerical tests of stochastic tomography, *Physics of the Earth and Planetary Interiors*, *67*(1-2), 180-193.

Wu, R. S., Z. Y. Xu, and X. P. Li (1994), Heterogeneity Spectrum and Scale-Anisotropy in the Upper Crust Revealed by the German Continental Deep-Drilling (Ktb) Holes, *Geophysical Research Letters*, *21*(10), 911-914.

Xu, W. B., C. Lithgow-Bertelloni, L. Stixrude, and J. Ritsema (2008), The effect of bulk composition and temperature on mantle seismic structure, *Earth and Planetary Science Letters*, *275*(1-2), 70-79.
21

# Appendix

## A.1 Correlation functions of mirrored random media: theoretical derivation

For a 2D isotropic Gaussian random medium, the perturbation correlation function and its PSDF can be expressed as (e.g., Zheng and Wu, 2008):

$$W(r_x, r_z) = \varepsilon^2 \exp\left(-\frac{r_x^2}{a^2}\right)\exp\left(-\frac{r_z^2}{a^2}\right), \tag{A1}$$

$$P(\kappa_x, \kappa_z) = \varepsilon^2 \pi a^2 \exp\left(-\frac{1}{4}\kappa_x^2 a^2\right) \exp\left(-\frac{1}{4}\kappa_z^2 a^2\right), \tag{A2}$$

where $r_x$ and $r_z$ is the horizontal and vertical correlation lag between two arbitrary perturbations; $\kappa_x$ and $\kappa_z$ are their corresponding wavenumbers; $\varepsilon$ is the perturbation strength; $a$ is the Gaussian correlation length that controls the perturbation scale. The 2D Gaussian correlation function can be decoupled by multiplication of two 1-D Gaussian functions, $\exp\left(-\frac{r_x^2}{a^2}\right)$ for the $x$-axis and $\exp\left(-\frac{r_z^2}{a^2}\right)$ for the $z$-axis. For a mirrored random medium with a symmetry axis at $z = 0$ (see Figure 2b), the correlation function along the $x$-axis is still a Gaussian function. However, the correlation function along the $z$-axis needs to be reconsidered.

To make the derivation easier to follow, we first consider the correlation function for a 1-D mirror medium. Assume we have a 1-D random medium (no mirror yet) with a limited thickness $L$. The velocity perturbation correlation function is:

$$\langle \delta v(z)\, \delta v(z+r) \rangle = \frac{1}{L}\int_0^L \delta v(z)\delta v(z+r)dz = W(r), \tag{A3}$$

$$W(r) = \varepsilon^2 \exp\left(-\frac{r^2}{a^2}\right), \tag{A4}$$

where $\langle\ \rangle$ implies the statistical ensemble averaging and an ensemble is defined as a set of the medium realizations; $\delta v = \frac{1}{2}\left(\frac{v_0^2}{v^2} - 1\right)$ is the fractional velocity fluctuation; $v_0$ and $v$ are the background and true velocity, respectively; $r$ is the vertical distance between two perturbations.

We can extend the random layer $\delta v(z)$ from $z \in [0\ L]$ to $z \in [-L\ 0]$ by a mirror projection (see Figure 2b). The mirrored random layer $\delta v_{mirror}(z)$ with a thickness of $2L$ would be expressed as:

$$\delta v_{mirror}(z) = \begin{cases} \delta v(z), z \in [0, L] \\ \delta v(-z), z \in [-L, 0] \end{cases}. \tag{A5}$$

Consequently, the 1D correlation functions $W_{mirror}(r)$ could be written as:

$$\begin{aligned}
W_{mirror}(r) &= \langle \delta v_{mirror}(z)\delta v_{mirror}(z+r)\rangle = \frac{1}{2L}\int_{-L}^{L}\langle \delta v_{mirror}(z)\delta v_{mirror}(z+r)dz\rangle \\
&= \frac{1}{2L}\left[\int_{-L}^{0}\langle \delta v(-z)\delta v(-z+r)dz\rangle + \int_{0}^{L}\langle \delta v(z)\delta v(z+r)dz\rangle\right] \\
&= \frac{1}{2L}\left[\int_{-L}^{0}\langle \delta v(-z)\delta v(-z+r)dz\rangle\right] + \frac{1}{2}W(r) \\
&= \frac{1}{2L}\left[\int_{0}^{L}\langle \delta v(z)\delta v(r-z)dz\rangle\right] + \frac{1}{2}W(r) \\
&= \frac{1}{2L}\int_{0}^{r}\langle \delta v(z)\delta v(r-z)dz\rangle + \frac{1}{2L}\int_{r}^{L}\langle \delta v(z)\delta v(r-z)dz\rangle + \frac{1}{2}W(r),
\end{aligned} \tag{A6}$$

where has three terms. The $z$ and $r - z$ in the first term in equation (A6) are $\geq 0$.

Equation (A6) can be calculated as:

$$\frac{1}{2L}\int_0^r \langle\delta(z)\delta(r-z)dz\rangle = \frac{1}{2L}\int_0^r \varepsilon^2 \exp(-\frac{(z-r+z)^2}{a^2})\,dz$$

$$= \frac{1}{2L}\varepsilon^2 \cdot \frac{1}{2}\sqrt{\pi}a\, erf\left(\frac{r}{a}\right), \quad (A7)$$

where $erf(x) = \frac{2}{\sqrt{\pi}}\int_0^x e^{-t^2}dt$ is the Gauss error function. This term is the correction of the correlation near the boundary ($z \in [0\ r]$) caused by the mirrored perturbations.

In the second term of equation (A6), the $z \geq 0$ while $r - z \leq 0$. The second term could be rewritten as:

$$\frac{1}{2L}\int_r^L \langle\delta v(z)\delta v(r-z)dz\rangle = \frac{1}{2L}\int_0^L \langle\delta v(z)\delta v(z-r)dz\rangle - \frac{1}{2L}\int_0^r \langle\delta v(z)\delta v(z-r)dz\rangle$$

$$= \frac{1}{2}B(r) - \frac{r}{2L}\varepsilon^2 \exp\left(-\frac{r^2}{a^2}\right). \quad (A8)$$

By inserting equations (A7) and (A8), equation (A6) can be rewritten as:

$$W_{mirror}(r) = W(r) + \frac{1}{2L}\varepsilon^2 \left[\frac{1}{2}\sqrt{\pi}a\, erf\left(\frac{r}{a}\right) - r\exp\left(-\frac{r^2}{a^2}\right)\right]. \quad (A9)$$

Here we can see the mirrored Gaussian random has a different correlation function compared to the one-layer Gaussian random. There are two terms to correct the correlation functions nearby the symmetric axis. In addition, these two correction terms are related to the random medium thickness $L$ as well as the correlation lag $r$. If the layer thickness $L$ is infinite, $W_{mirror}$ will be identical as $W$. If the correlation lag $r$ is comparable with $L$, the correction terms could not be ignored.

We can also get the PSDF for the mirrored medium using the Fourier transform:

$$P_{mirror}(\kappa_r) = \int W_{mirror}(r) e^{-i\kappa_r r} dr$$

$$= P(\kappa_r) + \frac{1}{2L}\varepsilon^2 \left\{ \frac{1}{2}\sqrt{\pi}a \, FT\left[\mathrm{erf}\left(\frac{r}{a}\right)\right] + \frac{1}{2}i\sqrt{\pi}a^3\kappa_r \exp\left(-\frac{1}{4}a^2\kappa_r^2\right) \right\}$$

$$= P(\kappa_r)$$

$$+ \frac{1}{2L}\varepsilon^2 \left\{ -\frac{1}{2}i\sqrt{\pi}a \, \frac{2\exp\left(-\frac{1}{4}\kappa_r^2 a^2\right)}{\kappa_r} + \frac{1}{2}i\sqrt{\pi}a^3\kappa_r \exp\left(-\frac{1}{4}a^2\kappa_r^2\right) \right\} \qquad (A10)$$

$$= P(\kappa_r) + \frac{i}{4L}\left(a^2\kappa_r - \frac{2}{\kappa_r}\right)\varepsilon^2\sqrt{\pi}a \exp\left(-\frac{1}{4}\kappa_r^2 a^2\right)$$

$$= \varepsilon^2\sqrt{\pi}a \exp\left(-\frac{1}{4}\kappa_r^2 a^2\right)\left[1 + \frac{i}{4L}\left(a^2\kappa_r - \frac{2}{\kappa_r}\right)\right],$$

where $\kappa_r$ is the wavenumber corresponding to the correlation lag $r$.

For 2-D case, the mirrored medium with a horizontal symmetry axis (such as Figure 2b) and the randomness is isotropic. We can obtain the correlation function and the associated spectrum in 2D:

$$W_{mirror}(r_x, r_z) = \varepsilon^2 \exp\left(-\frac{r_x^2}{a^2}\right)\left\{\exp\left(-\frac{r_z^2}{a^2}\right) + \frac{1}{2L}\left[\frac{1}{2}\sqrt{\pi}a\,\mathrm{erf}\left(\frac{r_z}{a}\right) - r_z \exp\left(-\frac{r_z^2}{a^2}\right)\right]\right\}, \qquad (A11)$$

$$P_{mirror}(\kappa_x, \kappa_z) = \varepsilon^2 \pi a^2 \exp\left(-\frac{1}{4}\kappa_x^2 a^2\right)\left[\exp\left(-\frac{1}{4}\kappa_z^2 a^2\right) + \frac{i}{4L}\left(a^2\kappa_z - \frac{2}{\kappa_z}\right)\right]. \qquad (A12)$$

## A.2 Numerical validation of the correlation function of mirrored random media for 1D case.

We validate the new correlation function for mirrored random perturbations by numerical experiments for 1D case. We first generate a 1D Gaussian random heterogeneous model in the spectral domain (e.g., Shapiro and Kneib, 1993), called one-layer Gaussian medium. In the spectral domain, for each wavenumber, we set its spectral amplitude according to a Gaussian function and assign it with a random phase within $[-\pi\ \pi]$. Then we transform the perturbations from the spectral domain to the spatial domain to form one random medium realization. Then we generate the mirrored Gaussian medium by mirror extending the one-layer Gaussian medium. To show the relationship between the correlation function of the mirrored Gaussian medium and the perturbation scale, we set four cases with Gaussian correlation length $a$ as 25 m, 50 m, 75 m and 100 m. The perturbation strength $\varepsilon$ in four case are the same at 0.01. The thickness of one-layer Gaussian medium is 500 m while the mirrored Gaussian random medium is 1000 m. After 100 random realizations, we can calculate the statistical ensemble averaged correlation functions and compare them with the theoretical ones from equations (A4) and (A9) (Figure 7). From Figure 7, we can see the new formula for the mirrored Gaussian medium fit measurements very well, among the correlation lag from 0 to a few times of $a$. For one-layer Gaussian medium, the correlation functions converge to zero after $2a$ while the mirrored Gaussian medium does not. The difference of correlation function between the one-layer Gaussian medium and the mirrored Gaussian medium becomes more obvious when the perturbation scale is large. There are some small discrepancies between the theoretical correlation functions and measured ones due to limited realization times. These numerical experiments validate our new formula of the perturbation correlation function for the mirrored Gaussian medium is accurate. We can use the new perturbation correlation function to calculate the TCFs of reflected waves.

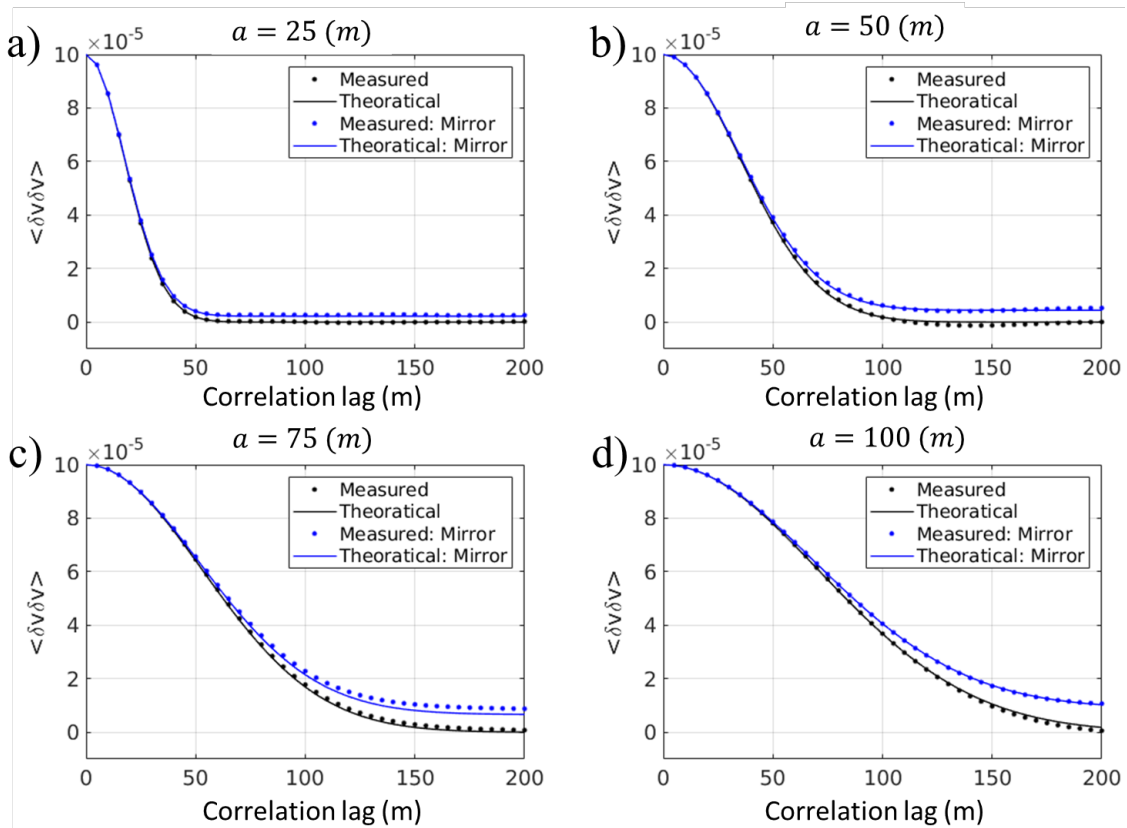

Figure 7. Comparison of the measured correlation functions using 100 random realizations and the theoretical ones. a), b) c) and d) are with different Gaussian correlation length $a$ from 25 m to 100 m. In each plot, the black dots are the measured correlation functions and black curves are the theoretical ones, for the one-layer Gaussian medium. The blue dots are the measured correlation functions and blue curves are their theoretical values, for the mirrored Gaussian medium.